\documentclass[aps,twocolumn,showpacs]{revtex4}
\usepackage{epsf}

\begin{document}

\title{\null\vspace*{-1truecm}  \hfill\mbox{\small IISc-CHEP-8/04}\\
            \vspace*{-0.2truecm}\hfill\mbox{\tt\small quant-ph/0405128}\\
\vspace*{0.5truecm}Quantum Random Walks do not need a Coin Toss}
\author{Apoorva Patel}
\email{adpatel@cts.iisc.ernet.in}
\altaffiliation[Also at ]{Supercomputer Education and Research Centre,
                      Indian Institute of Science, Bangalore-560012, India.}
\author{K.S. Raghunathan}
\email{ksraghu@yahoo.com}
\author{Pranaw Rungta}
\email{pranaw@cts.iisc.ernet.in}
\affiliation{Centre for High Energy Physics, Indian Institute of Science,
             Bangalore-560012, India}
\date{\today}

\begin{abstract}
\noindent
Classical randomized algorithms use a coin toss instruction to explore
different evolutionary branches of a problem. Quantum algorithms, on the
other hand, can explore multiple evolutionary branches by mere superposition
of states. Discrete quantum random walks, studied in the literature, have
nonetheless used both superposition and a quantum coin toss instruction.
This is not necessary, and a discrete quantum random walk without a quantum
coin toss instruction is defined and analyzed here. Our construction eliminates
quantum entanglement from the algorithm, and the results match those obtained
with a quantum coin toss instruction.
\end{abstract}
\maketitle

\section{Motivation}

Random walks are a fundamental ingredient of non-deterministic algorithms
\cite{motwani},
and are used to tackle a wide variety of problems---from graph structures
to Monte Carlo samplings. Such algorithms have many branches, which are
explored probabilistically, to estimate the correct result. A classical
computer can explore only one branch at a time, so typically the algorithm
is executed several times, and the estimate of the final result is extracted
from the ensemble of individual executions by methods of probability theory.
To ensure that different branches are explored in different executions, one
needs non-deterministic instructions, and they are provided in the form of
random numbers. A coin toss is the simplest example of a random number
generator, and it can be included as an instruction for a probabilistic
Turing machine.

A quantum computer can explore multiple branches in a different manner, i.e.
by using a superposition of states. The probabilistic result can then be
arrived at by interference of amplitudes corresponding to different branches.
Thus as long as the means to construct a variety of superposed states exist,
there is no a priori reason to include a coin toss as an instruction for
a (probabilistic) quantum Turing machine. This is obvious enough, and indeed
continuous time quantum random walks have been studied without recourse to a
coin toss instruction \cite{farhi}.
Nevertheless, a coin toss instruction has been considered necessary in
construction of discrete time quantum random walks (see for instance,
Refs.\cite{kempe,ambainis}).
In this article, we demonstrate that this is a misconception arising out of
unnecessarily restrictive assumptions. We explicitly construct a quantum
random walk on a line without using a coin toss instruction, and analyze
its properties.

There also exists confusion in the literature about different scaling
behavior of discrete and continuous time quantum random walk algorithms
(see again, Refs.\cite{kempe,ambainis}), because the former have
been constructed using a coin toss instruction while the latter do not
contain a coin toss instruction. Our work eliminates this confusion in
the sense that scaling behavior of discrete and continuous time quantum
random walk algorithms, both constructed without a coin toss instruction,
would coincide. Thereafter, a quantum coin would be an additional resource;
if its inclusion can improve scaling behavior of some quantum algorithms,
that should not be a surprise.

\section{Quantum Random Walk on a Line}

A random walk is a diffusion process, which is generated by the Laplacian
operator in the continuum. To construct a discrete quantum walk, we must
discretize this process using evolution operators that are both unitary
and ultra-local (an ultra-local operator vanishes outside a finite range
\cite{ultra}).
To begin with, consider the walk on a line. The allowed positions are
labeled by integers, and the simplest translation invariant ultra-local
discretization of the Laplacian operator is
\begin{equation}
H |n\rangle ~\propto~ \big[ -|n-1\rangle + 2|n\rangle - |n+1\rangle \big] ~.
\end{equation}
The corresponding evolution operator is
\begin{equation}
U(\Delta t) ~=~ \exp(iH\Delta t) ~=~ 1 + iH\Delta t + O((\Delta t)^2) ~.
\end{equation}
With a finite $\Delta t$, $U$ has an exponential tail and so it is not
ultra-local. The evolution operator can be made ultra-local by truncation,
say by dropping the $O((\Delta t)^2)$ part, but then it is not unitary.
One may search for ultra-local translationally invariant unitary evolution
operators using the ansatz
\begin{equation}
U |n\rangle ~=~ a|n-1\rangle + b|n\rangle + c|n+1\rangle ~,
\end{equation}
but then the orthogonality constraints between different rows of the unitary
matrix make two of $\{a,b,c\}$ vanish, and one obtains a directed walk
instead of a random walk.

One way to bypass this problem and construct a ultra-local unitary random walk
is to enlarge the Hilbert space and add a quantum coin toss instruction, e.g.
\begin{equation}
U = \sum_n \Big[
    |\!\uparrow\rangle\langle\uparrow\!|     \otimes |n+1 \rangle\langle n|
  + |\!\downarrow\rangle\langle\downarrow\!| \otimes |n-1 \rangle\langle n|
           \Big] .
\label{evolcoin}
\end{equation}
This modification however brings its own set of caveats. If the coin state
is measured at every time step (in other words, if the coin is classical),
one gets no improvement over the classical random walk. With a unitary coin
evolution operator, that entangles the coin state with the position state,
the quantum walk performs better than the classical walk in certain algorithms.
But in this case, the final results depend on the initial state of the coin,
because quantum evolution is reversible and not Markovian. For example, the
final state distribution of the quantum walk depends on whether the initial
coin state was $|\!\uparrow\rangle$, or $|\!\downarrow\rangle$, or some linear
combination thereof. To get around this initial coin state sensitivity,
further algorithmic modifications such as averaging over initial coin states,
or intermittent coin measurements, or use of multiple coins, have been
suggested, but they still leave a feeling of something to be desired.

\subsection{Getting Rid of the Coin}

The way out of the above conundrum is familiar to lattice field theorists
\cite{staggered}.
It has also been used to simulate quantum scattering with ultra-local
operators \cite{richardson},
and to construct quantum cellular automata \cite{meyer}.
In its simplest version, the Laplacian operator is decomposed into its
even/odd parts, $H = H_e + H_o$,
\begin{equation}
H \propto \left(\matrix{
    \cdots&\cdots  &\cdots  &\cdots  &\cdots  &\cdots  &\cdots  &\cdots \cr
    \cdots&\hfill-1&\hfill 2&\hfill-1&\hfill 0&\hfill 0&\hfill 0&\cdots \cr
    \cdots&\hfill 0&\hfill-1&\hfill 2&\hfill-1&\hfill 0&\hfill 0&\cdots \cr
    \cdots&\hfill 0&\hfill 0&\hfill-1&\hfill 2&\hfill-1&\hfill 0&\cdots \cr
    \cdots&\hfill 0&\hfill 0&\hfill 0&\hfill-1&\hfill 2&\hfill-1&\cdots \cr
    \cdots&\cdots  &\cdots  &\cdots  &\cdots  &\cdots  &\cdots  &\cdots \cr
    }\right) ,
\end{equation}
\begin{equation}
H_e \propto \left(\matrix{
    \cdots&\cdots  &\cdots  &\cdots  &\cdots  &\cdots  &\cdots  &\cdots \cr
    \cdots&\hfill-1&\hfill 1&\hfill 0&\hfill 0&\hfill 0&\hfill 0&\cdots \cr
    \cdots&\hfill 0&\hfill 0&\hfill 1&\hfill-1&\hfill 0&\hfill 0&\cdots \cr
    \cdots&\hfill 0&\hfill 0&\hfill-1&\hfill 1&\hfill 0&\hfill 0&\cdots \cr
    \cdots&\hfill 0&\hfill 0&\hfill 0&\hfill 0&\hfill 1&\hfill-1&\cdots \cr
    \cdots&\cdots  &\cdots  &\cdots  &\cdots  &\cdots  &\cdots  &\cdots \cr
    }\right) ,
\end{equation}
\begin{equation}
H_o \propto \left(\matrix{
    \cdots&\cdots  &\cdots  &\cdots  &\cdots  &\cdots  &\cdots  &\cdots \cr
    \cdots&\hfill 0&\hfill 1&\hfill-1&\hfill 0&\hfill 0&\hfill 0&\cdots \cr
    \cdots&\hfill 0&\hfill-1&\hfill 1&\hfill 0&\hfill 0&\hfill 0&\cdots \cr
    \cdots&\hfill 0&\hfill 0&\hfill 0&\hfill 1&\hfill-1&\hfill 0&\cdots \cr
    \cdots&\hfill 0&\hfill 0&\hfill 0&\hfill-1&\hfill 1&\hfill 0&\cdots \cr
    \cdots&\cdots  &\cdots  &\cdots  &\cdots  &\cdots  &\cdots  &\cdots \cr
    }\right) .
\end{equation}
The two parts, $H_e$ and $H_o$, are individually Hermitian. They are
block-diagonal with a constant $2\times2$ matrix, and so they can be
exponentiated while maintaining ultra-locality. The total evolution
operator can therefore be easily truncated, without giving up either
unitarity or ultra-locality,
\begin{eqnarray}
U(\Delta t) = e^{i(H_e+H_o)\Delta t} &=&
e^{iH_e\Delta t} e^{iH_o\Delta t} + O((\Delta t)^2) \\
&=& U_e(\Delta t) U_o(\Delta t) + O((\Delta t)^2) ~. \nonumber
\end{eqnarray}
The quantum random walk can now be generated using $U_e U_o$ as the
evolution operator for the amplitude distribution $\psi(n,t)$,
\begin{equation}
\psi(n,t) = [U_e U_o]^t \psi(n,0) ~,
\label{walkt}
\end{equation}
The fact that $U_e$ and $U_o$ do not commute with each other is enough
for the quantum random walk to explore all possible states. The price
paid for the above manipulation is that the evolution operator is
translationally invariant along the line in steps of 2, instead of 1.

The $2\times2$ matrix appearing in $H_e$ and $H_o$ is proportional to
$(1-\sigma_1)$, and so its exponential will be of the form $(c1+is\sigma_1)$,
$|c|^2+|s|^2=1$. A random walk should have at least two non-zero entries
in each row of the evolution operator. Even though our random walk treats
even and odd sites differently by construction, we can obtain an unbiased
random walk, by choosing the $2\times2$ blocks of $U_e$ and $U_o$ as
$\frac{1}{\sqrt2}{1\,i \choose i\,1}$. The discrete quantum random walk
then evolves the amplitude distribution according to
\begin{eqnarray}
U_o |n\rangle &=& {1 \over \sqrt{2}}\Big[ |n\rangle + i|n+(-1)^n\rangle \Big] , \\
U_e |n\rangle &=& {1 \over \sqrt{2}}\Big[ |n\rangle + i|n-(-1)^n\rangle \Big] ,
\end{eqnarray}
\begin{equation}
U_e U_o |n\rangle = {1 \over 2}\Big[
    i|n-1\rangle + |n\rangle + i|n+1\rangle - |n+2(-1)^n\rangle \Big] .
\label{walkstep}
\end{equation}

\subsection{Relation to the Walk with a Coin}

Our construction of discrete quantum random walk has exchanged the up/down
coin states for the even/odd site label. In the language of lattice field
theory, this strategy resembles staggered fermions \cite{staggered},
while that with a coin (or spin) is akin to Wilson fermions \cite{wilson}.
Indeed, an explicit relation between our construction and that with a coin
can be established. Let
\begin{equation}
\Psi(n,t) \equiv \left(\matrix{ \psi(2n,t) \cr \psi(2n+1,t) \cr}\right)
\end{equation}
describe the amplitude distribution in a two-component notation. Then
Eqs.(\ref{walkt},\ref{walkstep}) are equivalent to the evolution
\begin{equation}
\Psi(N,t) = [U C]^t \Psi(N,0) ~,~~
C = {1\over\sqrt{2}} \left(\matrix{ 1 & i \cr i & 1 \cr}\right) ~,
\end{equation}
\begin{equation}
U|N\rangle = {1\over\sqrt{2}}|N\rangle + {i\sigma_1 \over \sqrt{2}}\sum_\pm
{1\pm\sigma_3 \over 2}|N \mp 1\rangle ~.
\label{evolnocoin}
\end{equation}
Here, for clarity, we have denoted the basis states for $\Psi$ by $|N\rangle$.
The symmetric coin operator $C$ mixes the up/down components of $\Psi$. The
walk operator $U$ distributes the amplitude equally between remaining at the
same site and moving to the neighboring sites. The projection operators
$(1\pm\sigma_3)/2$ pick out the amplitude components that move forward and
backward. Finally, the operator $\sigma_1$ interchanges the up/down components
of $\Psi$, producing what Ref.\cite{gridsrch1} has called the flip-flop walk.

It is also instructive to note that while the diffusion operator $H$ has
the structure of a second derivative, its two parts, $H_e$ and $H_o$, have
the structure of a first derivative. This split is reminiscent of the
``square-root'' one takes to go from the Klein-Gordon operator to the
Dirac operator. For a quantum random walk with a coin, this feature has
been used to construct an efficient search algorithm on a spatial lattice
of dimension greater than one \cite{gridsrch1,gridsrch2}.
Reanalysis of that problem is in progress, without using a coin, in view
of our results \cite{inprogress}.

\section{Analysis of the Walk}

It is straightforward to analyze the properties of the walk in
Eq.(\ref{walkstep}) using the Fourier transform:
\begin{equation}
\widetilde\psi(k,t) = \sum_n e^{ikn} \psi(n,t) ~,~
\end{equation}
\begin{equation}
\psi(n,t) = \int_{-\pi}^{\pi} {dk \over 2\pi}~e^{-ikn} \widetilde\psi(k,t) ~.
\end{equation}
The evolution of the amplitude distribution in Fourier space is easily
obtained by splitting it into its even/odd parts:
\begin{equation}
\psi \equiv \left(\matrix{ \psi_e \cr \psi_o \cr}\right) ~,~~
\psi(k,t) = [M(k)]^t \psi(k,0) ~,~
\end{equation}
\begin{equation}
M(k) = \left(\matrix{ -ie^{ik}\sin k & i\cos k        \cr
                      i\cos k        & ie^{-ik}\sin k \cr} \right) ~.
\end{equation}
The unitary matrix $M$ has the eigenvalues,
$\lambda_\pm \equiv e^{\pm i\omega_k}$
(this $\pm$ sign label continues in all the results below),
\begin{equation}
\lambda_\pm = \sin^2 k \pm i\cos k \sqrt{1+\sin^2 k} ~,~~
\omega_k = \cos^{-1}(\sin^2 k) ~,
\end{equation}
with the (unnormalized) eigenvectors,
\begin{eqnarray}
e_\pm &\propto& \left(\matrix{ -\sin k \pm \sqrt{1+\sin^2 k} \cr 1 \cr} \right)
~, \nonumber\\
      &\propto& \left(\matrix{ 1 \cr \sin k \pm \sqrt{1+\sin^2 k} \cr} \right)
~.
\end{eqnarray}
The evolution of amplitude distribution then follows
\begin{equation}
\widetilde\psi(k,t) = e^{ iw_k t} \widetilde\psi_+(k,0)
                    + e^{-iw_k t} \widetilde\psi_-(k,0) ~,
\end{equation}
where $\widetilde\psi_\pm(k,0)$ are the projections of the initial amplitude
distribution along $e_\pm$. The amplitude distribution in the position space
is given by the inverse Fourier transform of $\widetilde\psi(k,t)$. While we
are unable to evaluate it exactly, many properties of the quantum random
walk can be extracted numerically as well as by suitable approximations.

Consider a walk starting at the origin, $\psi_{\rm o}(n,0)=\delta_{n,0}$.
Its amplitude distribution at later times is specified by
\begin{equation}
\widetilde\psi_{{\rm o},\pm}(k,0) = {\pm 1 \over 2\sqrt{1+\sin^2 k}}
     \left(\matrix{-\sin k \pm\sqrt{1+\sin^2 k} \cr 1 \cr}\right) ~,
\end{equation}
\begin{eqnarray}
\psi_{\rm o}(n,t) &=& {1 \over 2\pi}\int_{-\pi}^\pi
                  {dk~e^{-ikn} \over \sqrt{1+\sin^2 k}} \\
         &\times& \left(\matrix{ -i\sin\omega_k t \sin k + \cos\omega_k t
                  \sqrt{1+\sin^2 k} \cr i\sin\omega_k t \cr}\right) ~. \nonumber
\end{eqnarray}
This walk is asymmetric because our definitions treat even and odd sites
differently. We can get rid of the asymmetry by initializing the walk
as $\psi_{\rm s}(n,0)=(\delta_{n,0}+\delta_{n,1})/\sqrt{2}$. The walk is
then symmetric under $n\leftrightarrow(1-n)$, and the amplitude
distribution evolves according to
\begin{eqnarray}
\widetilde\psi_{{\rm s},\pm}(k,0) &=& {\pm 1 \over 2\sqrt{2(1+\sin^2 k)}} \\
     &\times& \left(\matrix{e^{ik} - \sin k \pm \sqrt{1+\sin^2 k} \cr
     1 + e^{ik}\sin k \pm e^{ik}\sqrt{1+\sin^2 k} \cr}\right) ~, \nonumber
\end{eqnarray}
\begin{eqnarray}
\label{symwalk}
&&\psi_{\rm s}(n,t) = {1 \over 2\pi}\int_{-\pi}^\pi
                  {dk~e^{-ikn} \over \sqrt{2(1+\sin^2 k)}} \\
&\times& \left(\matrix{  i\sin\omega_k t (e^{ik}-\sin k)
                        + \cos\omega_k t \sqrt{1+\sin^2 k} \cr
                         i\sin\omega_k t (1+e^{ik}\sin k)
                        + \cos\omega_k t~e^{ik}\sqrt{1+\sin^2 k} \cr}
         \right) ~. \nonumber
\end{eqnarray}
Figs.1-2 show the numerically evaluated probability distributions,
after $32$ time steps, for asymmetric and symmetric quantum random walks
respectively. Note that, by construction, the distributions after $t$
time steps remain within the interval $[-2t+1,2t]$.

\begin{figure}[tbh]
\epsfxsize=9truecm
\centerline{\epsfbox{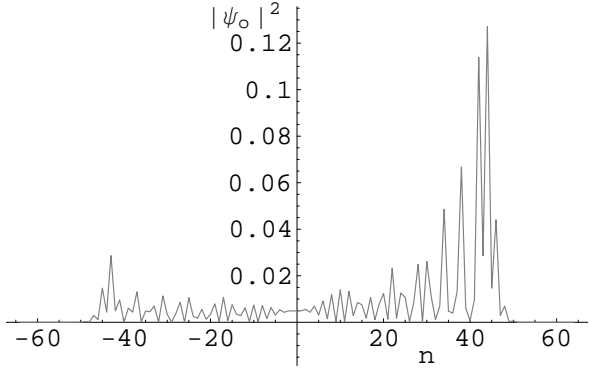}}
\vspace*{-1.5truecm}
\caption{Probability distribution after $32$ time steps for the asymmetric
quantum random walk $\psi_{\rm o}$.}
\end{figure}

\begin{figure}[tbh]
\epsfxsize=9truecm
\centerline{\epsfbox{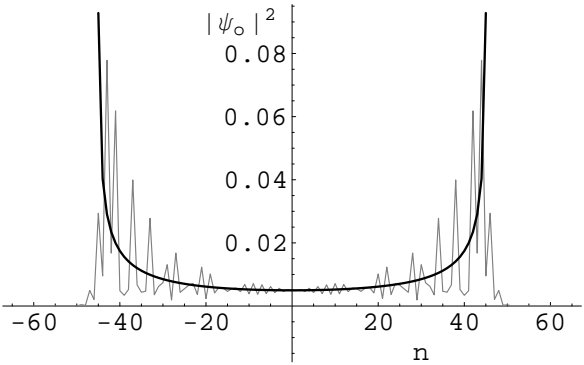}}
\vspace*{-1.5truecm}
\caption{Probability distribution after $32$ time steps for the symmetric
quantum random walk $\psi_{\rm s}$. The dark curve denotes the smoothed
distribution of Eq.(\ref{smooth}).}
\end{figure}

\subsection{Asymptotic Behavior of the Walk}

For large $t$, a good approximation to the distribution, Eq.(\ref{symwalk}),
can be obtained by the stationary phase method, as in Ref.\cite{nayak}.
The integral is periodic, and a sum of terms of the form
\begin{equation}
I(n,t) = \int_\pi^\pi {dk \over 2\pi}~g(k)~e^{i\phi(k,n,t)} ~.
\end{equation}
The highly oscillatory part of the integrand is determined by
$\phi(k,n,t) = -kn \pm \omega_k t$, while the remaining part $g(k)$
is bounded. Simple algebra yields
\begin{eqnarray}
{d  \omega_k \over dk  } &=& -{2\sin k \over \sqrt{1+\sin^2 k}} ~,\\
{d^2\omega_k \over dk^2} &=& -{2\cos k \over (1+\sin^2 k)^{3/2}} ~,\\
{d^3\omega_k \over dk^3} &=&  {4\sin k (1+\cos^2 k) \over (1+\sin^2 k)^{5/2}} ~.
\end{eqnarray}
The stationary point of the integral, $k=k_0$, has to satisfy
\begin{equation}
\alpha \equiv {n \over t} = \mp{2\sin k_0 \over \sqrt{1+\sin^2 k_0}} ~,
\end{equation}
which has a solution only for $n\in[-\sqrt{2}t,\sqrt{2}t]$.

We now separately consider the three cases:\\
(1) $|n| > \sqrt{2}t$: There is no stationary point in this case.
For $|n|=(\sqrt{2}+\epsilon)t$, $|d\phi(k)/dk|>\epsilon$, and repeated
integration by parts shows that the integral falls off faster than any
positive integer power of $\epsilon t$.\\
(2) $|n| = \sqrt{2}t$: In this case, there is a stationary point of order
2 at $k_0=\mp{\rm sgn}(n){\pi/2}$. The integral is therefore proportional
to $t^{-1/3}$. Explicitly,
\begin{eqnarray}
\psi_{\rm s}( \sqrt{2}t,t) &=& c~t^{-1/3}
     \left(\matrix{ (1+{1-i\over\sqrt{2}})\cos({\pi t \over \sqrt{2}}) \cr
                    (1-{1-i\over\sqrt{2}})\sin({\pi t \over \sqrt{2}}) \cr}
     \right) ,\\
\psi_{\rm s}(-\sqrt{2}t,t) &=& c~t^{-1/3}
     \left(\matrix{ (1-{1-i\over\sqrt{2}})\cos({\pi t \over \sqrt{2}}) \cr
                   (-1-{1-i\over\sqrt{2}})\sin({\pi t \over \sqrt{2}}) \cr}
     \right) ,\\
c &=& {1 \over 2\pi 3^{1/6}}~\Gamma\left({1\over3}\right)
       \approx 0.355 ~.
\end{eqnarray}
(3) $|n| < \sqrt{2}t$: There are two stationary points in this case,
$k_{01}\in(-\pi/2,\pi/2)$ and $k_{02}=\pi-k_{01}$, with
\begin{eqnarray}
\sin k_0 &=& \mp {n \over \sqrt{4t^2 - n^2}} ~,~ \\
\left|{d^2\omega_k \over dk^2}\right|_{k=k_0}
         &=& {\sqrt{4t^2-2n^2}~(4t^2-n^2) \over 4t^3} ~.
\end{eqnarray}
The integral is therefore proportional to $t^{-1/2}$. In terms of the phase,
\begin{equation}
\phi_0 = -k_{01} n + \omega_{k_0} t - (\pi/4) ~,
\end{equation}
the distribution amplitude is
\begin{eqnarray}
\psi_{\rm s} &=& {1 \over \sqrt{t}(4t^2-2n^2)^{1/4}} \nonumber\\
        &\times& \left[ \cos\phi_0 \left(\matrix{
                 ((1-i)n + 2t)/\sqrt{4t^2-n^2} \cr
                 \sqrt{4t^2-2n^2}/(2t+n) \cr}\right) \right. \\
             & & \left. + i\sin\phi_0 \left(\matrix{
                 \sqrt{4t^2-2n^2}/\sqrt{4t^2-n^2} \cr
                 ((1-i)n + 2t)/(2t+n) \cr}\right) \right] ~. \nonumber
\end{eqnarray}
The smoothed probability distribution, obtained by replacing the highly
oscillatory terms by their mean values, is
\begin{equation}
\label{smooth}
|\psi_{\rm s}|_{\rm smooth}^2 = {4t^2 \over \pi\sqrt{4t^2-2n^2}~(4t^2-n^2)} ~.
\end{equation}
(Here, the $n\leftrightarrow(1-n)$ symmetry can be restored by replacing
$n$ by $(n-{1\over2})$.)
As shown in Fig.2, it represents the average behavior of the distribution
very well. Its low order moments are easily calculated to be,
\begin{eqnarray}
\label{moment0}
\int_{n=-\sqrt{2}t}^{\sqrt{2}t} |\psi_{\rm s}|_{\rm smooth}^2 dn &=& 1 ~, \\
\int_{n=-\sqrt{2}t}^{\sqrt{2}t} |n| \cdot |\psi_{\rm s}|_{\rm smooth}^2 dn &=& t ~, \\
\int_{n=-\sqrt{2}t}^{\sqrt{2}t} n^2 |\psi_{\rm s}|_{\rm smooth}^2 dn
     &=& 2(2-\sqrt{2})t^2 ~.
\label{moment2}
\end{eqnarray}

The following properties of the quantum random walk are easily deduced
from all the above results:\\
$\bullet$ 
The probability distribution is double-peaked with maxima approximately at
$\pm\sqrt{2}t$. The distribution falls off steeply beyond the peaks, while
it is rather flat in the region between the peaks. With increasing $t$,
the peaks become more pronounced, because the height of the peaks decreases
slower than that for the flat region.\\
$\bullet$
The size of the tail of the amplitude distribution is limited by
$(\epsilon t)^{-1} \sim t^{-1/3}$, which gives
$\Delta n_> = \Delta(\epsilon t) = O(t^{1/3})$.
On the inner side, the width of the peaks is governed by
$|\omega_k^{''} t|^{-1/2} \sim t^{-1/3}$. For $|n|=(\sqrt{2}-\delta)t$,
this gives $\Delta n_< = \Delta(\delta t) = O(t^{1/3})$.
The peaks therefore make a negligible contribution to the probability
distribution, $O(t^{-1/3})$.\\
$\bullet$
Rapid oscillations contribute to the probability distribution (and hence
to its moments) only at subleading order. They can be safely ignored in
an asymptotic analysis, retaining only the smooth part of the probability
distribution.\\
$\bullet$
The quantum random walk spreads linearly in time, with a speed smaller
by a factor of $\sqrt{2}$ compared to a directed walk. This speed is a
measure of its mixing behavior and hitting probability. The probability
distribution is qualitatively similar to a uniform distribution over the
interval $[-\sqrt{2}t,\sqrt{2}t]$. In particular, $m^{th}$ moment of the
probability distribution is proportional to $t^m$.

These properties agree with those obtained in Ref.\cite{nayak}
for a quantum random walk with a coin-toss instruction, demonstrating that
the coin offers no advantage in this particular set up. (Extra factors of
$2$ appear in our results, because a single step of our walk is a product
of two nearest neighbor operators, $U_e$ and $U_o$.) It is important to
note that the properties of the quantum random walk are in sharp contrast
to those of the classical random walk. The classical random walk produces
a binomial probability distribution, which in the symmetric case has a
single peak centered at the origin and variance equal to $t$.

\subsection{The Walk in Presence of an Absorbing Wall}

The escape probability of the quantum random walk can be calculated by
introducing an absorbing wall, say between $n=0$ and $n=-1$. Mathematically,
the absorbing wall can be represented by a projection operator for $n\ge0$.
The unabsorbed part of the walk is given by
\begin{eqnarray}
\psi(n,t+1) &=& P_{n\ge0}U_e U_o~\psi(n,t) ~,\\
            &=& U_e U_o~\psi(n,t) - {1\over2}\delta_{n,-1}(i\psi(0,t)-\psi(1,t)) ~,
\nonumber
\end{eqnarray}
with the absorption probability,
\begin{equation}
P_{\rm abs}(t) = 1 - \sum_{n\ge0} |\psi(n,t)|^2 ~.
\end{equation}
Fig.3 shows the numerically evaluated probability distribution, in presence
of this absorbing wall, after $32$ time steps and with the symmetric initial
state. Comparison with Fig.2 shows that the absorbing wall disturbs the
evolution of the walk only marginally. The probability distribution in the
region close to $n=0$ is depleted as anticipated, while it is a bit of a
surprise that the peak height near $n=\sqrt{2}t$ increases slightly. As a
result, the escape speed from the wall is little higher than the spreading
speed without the wall. As shown in Fig.4, we find that the first two time
steps dominate absorption, $P_{\rm s,abs}(t=1)=0.25$ and $P_{\rm s,abs}(t=2)
=0.375$, with very little absorption later on. Asymptotically, the net
absorption probability approaches $P_{\rm s,abs}(\infty) \approx 0.4098$ for
the symmetric walk. (We also find, for the asymmetric walk starting at the
origin, $P_{\rm o,abs}(\infty)\approx0.2732$.) This value is smaller than
the corresponding result $P_{\rm abs}(\infty)=2/\pi$ for the symmetric
quantum random walk with a coin-toss instruction \cite{watrous}.

Thus the part of quantum random walk going away from the absorbing wall just
takes off at a constant speed, hardly ever returning to the starting point.
Again, this behavior is in a sharp contrast to that of the classical random
walk. A classical random walk always returns to the starting point, sooner
or later, and so its absorption probability approaches unity as
$t\rightarrow\infty$.

\begin{figure}[tbh]
\epsfxsize=9truecm
\centerline{\epsfbox{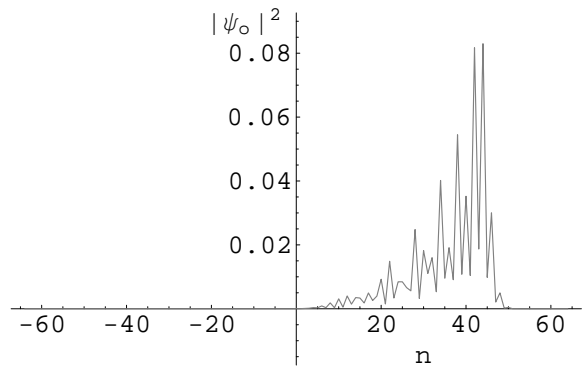}}
\vspace*{-1.5truecm}
\caption{Probability distribution after $32$ time steps for the symmetric
quantum random walk, with an absorbing wall on the left side of $n=0$.}
\end{figure}

\begin{figure}[tbh]
\vspace*{-0.5truecm}
\epsfxsize=9truecm
\centerline{\epsfbox{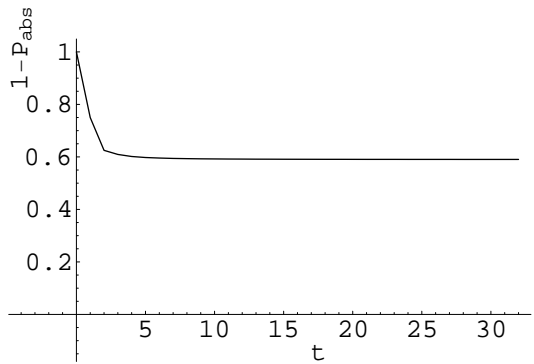}}
\vspace*{-1.4truecm}
\caption{Time dependence of the probability for the symmetric quantum
random walk to remain unabsorbed, in presence of an absorbing wall to
the left of $n=0$. The walk gets within a few per cent of the asymptotic
escape probability in just two time steps.}
\end{figure}

\subsection{Comparison to the Walk with a Coin}

The above results bring out the differences of our quantum random
walk construction compared to that of Refs.\cite{nayak,watrous}:\\
(1) We have absorbed the two states of the coin in to the even/odd site
label at no extra cost. This is possible because, due to its discrete
symmetry, the walk with a coin effectively uses only half the sites (e.g.
for a walk starting at origin, the amplitude distribution is restricted to
only odd sites at odd $t$ and only even sites at even $t$). Our walk makes
use of all the sites at every instance.\\
(2) It can be seen from Eqs.(\ref{evolnocoin},\ref{evolcoin}) that,
at every time step, $\Psi$ has 50\% probability to stay put at the same
location, while the walk with a coin has no probability to remain at the
same location. Yet, both achieve the same spread of amplitude distribution,
as exemplified by the moments in Eqs.(\ref{moment0}-\ref{moment2}). This
means that our walk is smoother---more directed and less zigzag.\\
(3) When the coin is considered a separate degree of freedom, quantum
evolution entangles the coin and the walk position. On the other hand,
when the coin states are made part of the position space, as we have done,
entanglement disappears completely---only superposition representing the
amplitude distribution survives \cite{entangle}.
This elimination of quantum entanglement would be a tremendous advantage in
any practical implementation of the quantum random walk, because quantum
entanglement is highly fragile against environmental disturbances while
mere superposition is much more stable. The cost for gaining this advantage
is the loss of short distance homogeneity---translational invariance holds
in steps of 2 instead of 1.

\section{Extensions and Outlook}

The quantum random walk on a line is easily converted to that on a circle
by imposing periodic boundary conditions. When the circle has $N$ points,
the only change in the analysis is to replace the integral over $k$ in the
inverse Fourier transform by a discrete sum over $k$-values separated by
$2\pi/N$. Since the quantum random walk spreads essentially uniformly,
there is not much change in its behavior. All that one has to bear in
mind is that, on a long time scale, unitary evolution makes the walk cycle
through phases of spreading out and contracting towards the initial state.

Going beyond one dimension, the strategy of constructing discrete
ultra-local unitary evolution operators by splitting the Hamiltonian in to
block-diagonal parts is applicable to random walks on any finite-color
graph \cite{richardson}.
One just constructs $2\times2$ block unitary matrices for each color of the
graph, and the single time step evolution operator becomes the product of 
all the block unitary matrices. In particular, the $d-$dimensional hypercubic
walk can be constructed as a tensor product of $d$ one-dimensional walks,
using $2d$ block unitary matrices.

Our results clearly demonstrate that discrete quantum random walks with
useful properties can be constructed without a coin toss instruction.
The addition of a coin toss instruction may still be beneficial in specific
quantum problems. A coin is an extra resource, and there are known instances
where the addition of a coin toss instruction makes classical randomized
algorithms have a better scaling behavior compared to their deterministic
counterparts \cite{motwani}.
One may hope for a similar situation in the quantum case too, keeping in
mind that a careful initialization of the quantum coin state would be a
must in such cases.

A clear advantage of quantum random walks is their linear spread in time,
compared to square-root spread in time for classical random walks.
So they are expected to be useful in problems requiring fast hitting times.
Some examples of this nature have been explored in graph theoretical and
sampling problems (see Refs.\cite{kempe,ambainis} for reviews),
and more applications need to be investigated.


\begin{thebibliography}{99}
\bibitem{ultra}{We deliberately use the label ``ultra-local'' here,
because ''local'' physical interactions (especially when described
in the renormalization group framework) include all those that fall
off exponentially or faster with distance.}
\bibitem{entangle}{Entanglement is always defined with respect to a specific
division of the whole system in to its parts. If the division scheme is
altered, entanglement can change.}
\end{thebibliography}
\end{document}